\documentclass[aps,epsf,eqsecnum,feynmp,graphics,amsfonts, amssymb, latexsym, amsmath]{revtex4}
\usepackage[latin1]{inputenc}%
\usepackage{graphics}
\usepackage{showlabels}
\usepackage{amsfonts, amssymb, latexsym, amsmath}
\def\aut#1{#1}
\def\comment#1{}
\begin{document}
\title{Dependence of Variational Perturbation Expansions
 on Strong-Coupling Behavior.
Inapplicability of $ \delta $-Expansion to Field Theory.}
\author{B.~Hamprecht and H.~Kleinert}
\address{Institut für Theoretische Physik, Freie Universität Berlin,\\
Arnimallee 14, D-14195 Berlin, Germany\\
{\scriptsize e-mails: bodo.hamprecht@physik.fu-berlin.de}
{\scriptsize e-mails: hagen.kleinert@physik.fu-berlin.de}}
\begin{abstract}
We show that in applications of variational
theory to quantum field theory
it is essential to account for the
correct Wegner
exponent $ \omega $ governing  the  approach to the
strong-coupling,
or scaling limit.
Otherwise
the
procedure either does not converge
at all or
 to the
wrong limit.
This invalidates
all papers
applying the
so-called $ \delta $-expansion
to quantum field theory.
\end{abstract}
\maketitle
\section{Introduction}
Variational
perturbation theory is a powerful tool for extracting
non-perturbative strong-coupling results from weak-coupling expansions.
It was initially invented in  quantum mechanics
as a re-expansion of the
perturbation series
of the
action \cite{refs}
\begin{eqnarray}
{\cal A}=\int_{t_a}^{t_b} dt\left[\frac{M}{2}\dot x^2-\frac{ \omega ^2}{2}x^2-V^{\rm int}(x)\right] ,
\label{@}\end{eqnarray}
which arises from splitting the potential
into a quadratic part
$V_{\Omega}^{(0)}\equiv \Omega ^2x^2/2$, with an arbitrary
trial frequency $ \Omega $,
and
 an interacting part
\begin{equation}
V_{\Omega}^{\rm int}\equiv \delta \left[
\frac{( \omega ^2- \Omega ^2)}{2}x^2+V^{\rm int}(x)\right] .
\label{@varint}\end{equation}
\comment{
\begin{equation}
{\cal A}=
{\cal A}_0
+
{\cal A}^{\rm int},~~~~~~
{\cal A}_0=\int_{t_a}^{t_b} dt
\left[\frac{M}{2}\dot x^2-\frac{ \Omega ^2}{2}x^2\right] ,~~~~
{\cal A}^{\rm int}=
\int_{t_a}^{t_b} dt\,V^{\rm int}_{\Omega}\equiv
\int_{t_a}^{t_b} dt\left[
-\frac{( \omega ^2- \Omega ^2)}{2}x^2-V^{\rm int}(x)\right] ,
\label{@}\end{equation}
}
The perturbation expansion is then performed in powers of $ \delta $, setting
 $ \delta =1$ at the
end,
and
  optimizing the
result in
$ \Omega $
guided by
 the
{\em principle of minimal sensitivity\/} \cite{Stevenson}.
The history and
 convergence properties
are discussed in the
textbook \cite{Hagen}.
Due
to the
prefactor
$ \delta $ in (\ref{@varint}),
the
procedure is often called $ \delta $-expansion \cite{refs}.
For the
anharmonic oscillator,
convergence was proved to be exponentially fast
for finite  \cite{finiteg}
as well as for infinite coupling strength
\cite{KleinertJanke,Guida,Hagen}.

In recent years, the
method has been extended
in a simple but essential way to allow for the
resummation of divergent
perturbation expansions
in quantum field-theories
 \cite{strong,epsilon}.
The most important new feature
of this {\em field-theoretic
variational perturbation theory\/}
is that it
accounts for
the
anomalous power
 approach
to the
strong-coupling limit
 which the
$ \delta $-expansion
cannot do.
This approach is governed by an irrational
critical exponent
$ \omega $
as was first shown
 by Wegner \cite{Wegner}
in the
context of critical
phenomena.
In contrast to the
$ \delta $-expansion, the
field-theoretic
variational perturbation
expansions  {\em cannot\/} be derived from the
action by
adding and
 subtracting a harmonic term
as in (\ref{@varint}).
The new theory has led to the
so-far most accurate determination of critical
exponents
via quantum field theory, as  amply demonstrated
in the
textbook \cite{Verena}.
In particular, the
theory has
perfectly  explained
 the
experimentally best
known critical exponent
$ \alpha $ of the
specific heat of the
$ \lambda $-transition
measured in a satellite orbiting around the
earth
\cite{Lipa}.


In spite of the
existence of this
reliable quantum-field-theoretic
variational
perturbation theory,
the literature keeps
offering
applications of the
above quantum-mechanical
$ \delta $-expansion to quantum field theory,
for instance in  recent papers by
 Braaten and
 Radescu (BR) \cite{Braaten,BraatenBE}
and
 Ramos \cite{Ramos} (see also
\cite{fermions}).

It is the
purpose of this paper
to show what goes wrong
with such unjustified applications, and
 how the
proper quantum
field-theoretic variational perturbation theory
corrects the mistakes.

\section{Review of the
Method}
 Suppose, the
function $f(g)$ is given by a
 divergent series expansion around the
point $g=0$:
\begin{align}
\label{WEAK}
f_L(g) = \sum_{l=0}^L\; a_l\;g^l \,,
\end{align}
typically with factorial growth of the
coefficients $a_l$.
Suppose furthermore, that the
expected
 leading behavior of
$f(g)$ for large $g$
has the
general power structure:
\begin{align}
\label{STRONG}
f_M(g) = g^\alpha\;\sum_{m=0}^M\; b_m\;g^{-\omega\,m} \,,
\end{align}
where $ \omega $ is the
Wegner exponent of approach to the
strong-coupling limit.
In quantum mechanics, this exponent is
easily  found
from the
naive scaling
properties of the
action.
In quantum field theory, however,
it is an initially unknown
number which has to be determined
from
the
above weak-coupling expansion
by a procedure to be
called {\em dynamical determination\/} of $ \omega $.

Assuming for a moment that this has been done,
the
$L$th order approximation to the
leading coefficient $b_0$ is given by \cite{Hagen}:
\begin{align}
\label{B0}
b_0^{(L)}(z) = z^{-\alpha}\;\sum_{l=0}^L\; a_l\;z^l \binom{L-l+(l-\alpha)/\omega}{L-l} \,,
\end{align}
where
the
$z\equiv g/ \Omega ^{1/\alpha}$ is
the variational parameter to be optimized
for minimal
sensitivity on $z$.
A short reminder of
the derivation of this formula is given in
Appendix A.
An application to a
simple known function is shown in Appendix B.
For a
successful application
to  the
quantum-mechanical anharmonic oscillators,
the reader
is referred to
the
textbook \cite{Hagen}.
The exponent $ \omega $ is equal to $2/3$
for an $x^4$-anharmonic oscillator, and
 the
exponentially fast
convergence has a
last term decreasing like
$e^{-{\rm const}\times L^{1- \omega }}$.
 For the oscillator,
the number $ \omega $ is found directly
from the dimensional
analysis
in
Appendix A.
As mentioned above,
such an analysis will not be applicable
in quantum field theory, where $ \omega $
will be anomalous and
must be determined dynamically.

Most often we want
to calculate
a quantity $f(g)$
which goes to a constant in the
strong-coupling limit
$f(g)\displaystyle\longrightarrow f^*$.
This is the
case for all
critical exponents.
Then we must set
$ \alpha =0$ in (\ref{STRONG}) and
 (\ref{B0}),
which implies that for infinite $g$:
\begin{align}
\label{betaV}  \beta (g)=
\left.\frac{d~\log f(g)}{d~\log~g}\right|_{g\rightarrow \infty}= 0~.
\end{align}
If $ \beta (g)$ is reexpressed
as a function of $f$, this implies
$\beta (f^*)=0$,
the
standard requirement for the
 existence of a critical point
in quantum field theory if $f(g)=g_R(g)$ is the
renormalized coupling strength
 as a function of the
bare coupling strength $g$.

The dynamical
determination of $ \omega $ proceeds now by
treating not only $f(g)$ but also the
beta function (\ref{betaV})
 according to the
rules of variational perturbation theory,
and
 determining  $\omega$
to make $ \beta ^*= \beta (\infty)$
vanish, which is done by
 optimizing
 the
equation
of $z$

\begin{align}
\label{LOG}
\sum_{l=0}^L\;  \beta _l\;z^l \binom{L-l+l/\omega}{L-l}=0 ,
\end{align}
where
$ \beta _l$ are the
coefficients of the
expansion   of
 (\ref{betaV}) with respect to $g$.
Minimal sensitivity is reached
 for
a vanishing
 derivative with respect to $z$:
\begin{align}
\label{LOG1}
\sum_{l=1}^L\; \beta_l\;l\,z^{l-1} \binom{L-l+l/\omega}{L-l}=0 \,,
\end{align}
so that $z$ and
  $\omega$ are to be found as simultaneous
solutions of (\ref{LOG}) and
 (\ref{LOG1}).
\\
\section{Anomalous Dimensions}
As mentioned above,
a number of  authors have applied the
$ \delta $-expansion
to field theories. Most recently, this was done
for the
purpose
of calculating the
shift of the
critical temperature
 in a Bose-Einstein condensate caused by a small interaction
\cite{BraatenBE,Ramos}.
Since the
perturbation expansion for this quantity
is a function of $g/\mu$ where $\mu$ is the
chemical potential
which goes to zero
at the critical point,
we are faced with a
typical strong-coupling problem
of critical phenomena.
In order to justify the
application of the
$ \delta $-expansion
to this problem,
 BR \cite{Braaten}
studied the
convergence
properties
of the
 method
by applying it to
a certain  amplitude $ \Delta (g)$ of an
$O(N)$-symmetric $\phi^4$-field theory
in the
limit of large $N$, where
the model is exactly solvable.

Their procedure can be criticized in two ways.
First, the amplitude $ \Delta (g)$ they considered
 is not a good candidate for
a  resummation  by a $ \delta $-expansion
since
 it does not possess the
characteristic
strong-coupling power structure
 (\ref{STRONG})
 of quantum mechanics and
 field theory,
which the
final
resummed expression will always have.
The power structure is disturbed
by additional logarithmic terms.
Second, the
 $ \delta $-expansion
is equivalent to  choosing, on dimensional grounds, the
 exponent
 $\omega=2$ in (\ref{STRONG}), which is far from the
an approximate optimal value 0.843
to be derived below.
Thus the $ \delta $-expansion is
inapplicable, and this
 explains  the
 problems into which BR run in
their resummation attempt.
Most importantly, they
do
 not
 find
a well-shaped plateau of the variational expressions
 $ \Delta^{(L)} (g,z)$ as a function of $z$ which would be
 necessary for invoking the
principle of minimal sensitivity.
Instead, they observe
that the
zeros of the
first derivatives
$\partial _z \Delta^{(L)} (g,z)$
 run  away
far into the
complex plain.
Choosing the complex
 solutions
to determine their final resummed value misses the
correct one by 3\%
up to the 35th order.

One may improve the situation
by
 trying out
various different  $\omega$-values and
 choosing
the
best of them yielding
an acceptable plateau in $ \Delta (g,z)$.
This happens for $\omega \approx 0.843$.
However, even
for this optimal value, the
resummation result
never converges to the
correct limit.
For $ \Delta (g)$
the error happens to be numerically small, only
 0.1\%,
but it will be uncontrolled
in physical problems
where the result is unknown.

Let us explain these points in more detail.
BR
consider
the
weak-coupling series
with the reexpansion parameter $ \delta $:
\begin{align}
\label{BRAA1}
\Delta(\delta, g) &=-\sum_{l=1}^\infty
\Big(-\frac{\delta \,g}{\sqrt{1-\delta}}\Big)^l a_l~,  &\!\!\!\!\!\!
{\rm where}~~~
a_l &\equiv  \int_0^\infty K(x)f^l(x)\,dx\,,
\end{align}
with
\begin{align}
\label{BRAA1.1}
K(x) &\equiv  \frac{4 x^2}{\pi (1+x^2)^2}~,  &
f(x) &\equiv \frac{2}{x}\arctan \frac{x}{2}\,.
\end{align}
The geometric series in (\ref{BRAA1}) can be summed exactly,
and
 the
result may formally be reexpanded into
a strong-coupling series in $h\equiv
{\sqrt{1-\delta}}/({\delta \,g})$:
\begin{align}
\label{BRAA2}
\Delta(\delta, g) &= \int_0^\infty K(x) \frac{\delta g f(x)}{\sqrt{1-\delta}+\delta g f(x)}\,dx
=\sum_{m=0}^\infty b_m
\left(-h\right)^m,
 &  \mbox{where~~  }
b_m= \int_0^\infty K(x)f^{-m}(x)\,dx\,.
\end{align}
The strong-coupling limit  is found for $h\rightarrow 0$
where
$ \Delta \rightarrow b_0
=\int_0^\infty dx\, K(x)=1$.
The approach to this limit is, however,
{\em not\/}  given by a
strong-coupling expansion of the
form
 (\ref{BRAA2}). This would only happen
if all the
integrals $b_m$ were to exist which,
unfortunately, is not the
case
since
all
 integrals for $b_m$
with
  $m>0$
diverge at the
upper limit, where
\begin{align}
\label{BRAA1.2}
f(x) &= \frac{2}{x}\arctan \frac{x}{2} \sim \frac{\pi}{x}\,.
\end{align}
The exact behavior of $\Delta$
in the strong-coupling limit
 $h \to 0$
is found
 by studying the
effect of the
asymptotic  $\pi /x$-contribution of $f(x)$
to the
integral in (\ref{BRAA2}). For $f(x)=\pi /x$ we obtain
\begin{align}
\label{BRAA3}
\int_0^\infty K(x) \frac{1}{1+h / f(x)}\,dx = \frac{\pi^4+2\pi²h-\pi²h²+2h³+4\pi²h \log{h/\pi}}{(\pi²+h²)²}\,.
\end{align}
The logarithm of $h$
shows a mismatch with
(\ref{STRONG})
and prevents
 the
expansion
(\ref{BRAA1}) to be a candidate
for variational perturbation theory.

We now explain the
second criticism.
Suppose we
 ignore the
just-demonstrated fundamental
 obstacle and
follow the
rules of the
$\delta$-expansion, defining
 the
$L$th order
approximant  $\Delta( \delta ,\infty)$
by expanding (\ref{BRAA1})
 in powers of $\delta$ up to order $ \delta ^L$,
 setting $\delta=1$, and
 defining $z\equiv g$.
Then we
 obtain the
$L$th variational expression for $b_0$:
\begin{align}
\label{BRAAST}
b_0^{(L)}( \omega ,z)=\sum_{l=1}^L a_l z^l \binom{L-l+l/\omega}{L-l}~,
\end{align}
with $ \omega =2$, to be optimized in $z$.
This $ \omega $-value
would only be adequate
if the
approach to the
strong-coupling limit
 behaved like $A + B/h^2+\dots$, rather
than
 (\ref{BRAA3}).
This is the
reason why BR  find
no real regime of minimal sensitivity on $z$.

\begin{figure}[htp]
\begin{center}
\setlength{\unitlength}{1cm}
\begin{picture}(12,7)
\scalebox{1.}[1.]{\includegraphics*{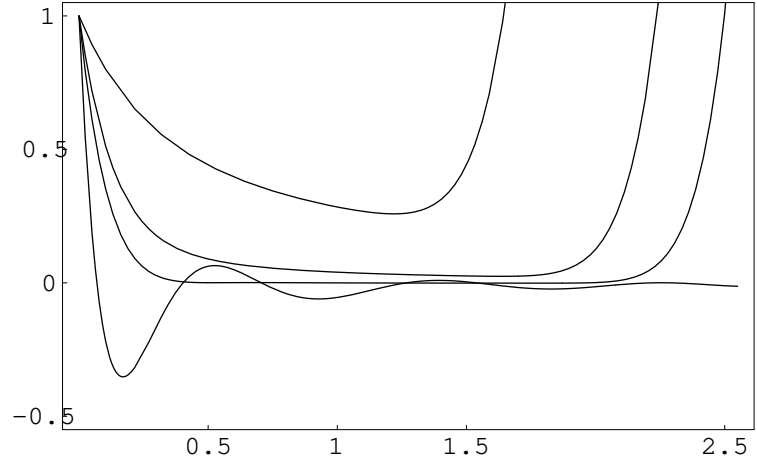}}
\end{picture}
\caption[BRI]{Plot of $1-b_0^{(L)}(\omega,z)$
versus $z$ for $L=10$ and
$\omega=0.6,~0.843,~ 1,~2~$.
The
curve
 with $\omega=0.6$
 shows oscillations. They decrease with
increasing $\omega$
and
 becomes flat
 at about
$\omega=0.843$. Further increase
of $ \omega $  tilts the
plateau and
 shows no regime of
minimal sensitivity.
At the
same time, the
minimum of the
curve rises
 rapidly above the
correct value of $1-b_0=0$,
as can be seen from the
upper two curves for $\omega=1$
and
  $\omega=2$, respectively. }
\label{BRI}
\end{center}
\end{figure}

Let us attempt to
improve the situation
by determining $\omega$ dynamically from equation (\ref{betaV}).
The result is
 $\omega \approx 0.843$, quite far from the
naive value $2$.
This value can also  be
estimated by inspecting plots of $\Delta^{(L)}(\omega,h)$
versus $h$ for several different $\omega$-values in
Fig. \ref{BRI}, and
 selecting the
one producing minimal sensitivity.
\begin{figure}[htp!]
\begin{center}
\setlength{\unitlength}{1cm}
\begin{picture}(19,11)
\scalebox{1.75}[1.75]{\includegraphics*{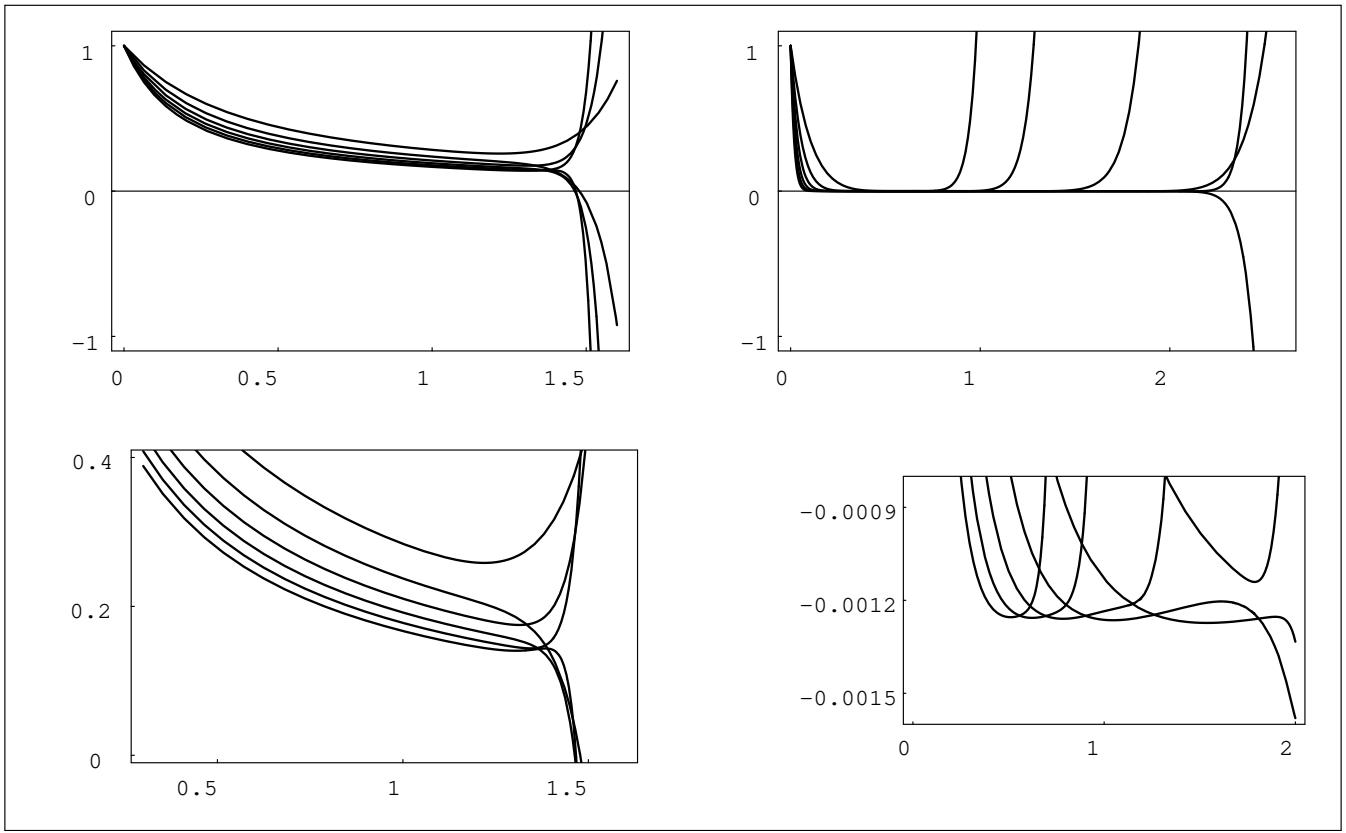}}
\end{picture}
\caption[BRII]{
Left-hand
 column shows plots
of
$1-b_0^{(L)}(\omega,z)$ for $L=10, ~17, ~24, ~31, ~38,~45$
with $\omega=~2$ of $ \delta $-expansion
of BR,
right-hand
 column with optimal $\omega=0.843$.
The lower row  enlarges the
interesting plateau regions
of the
plots above.
Only the
right-hand
 side
shows minimal sensitivity, and
 the
associated
plateau lies closer
to the
correct value  $1-b_0 = 0$
than
 the
minima in
  the
left column by
 two orders of magnitude.
Still the
right-hand
 curves do not approach the
exact limit
for $L\rightarrow \infty$
due to the
wrong strong-coupling behavior of the
initial function.}
\label{BRII}
\end{center}
\end{figure}%
\begin{figure}[htp!]
\begin{center}
\setlength{\unitlength}{1cm}
\begin{picture}(19,6)
\scalebox{1.}[1.]{\includegraphics*{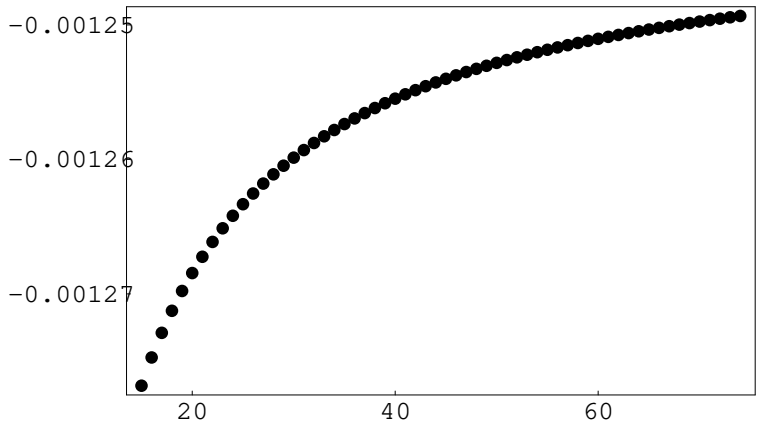}}
\end{picture}
\caption[BRIII]{Deviation of $1-b^{(L)}_{0,{\rm plateau}}(\omega=0.843)$
from zero
 as a function of the
order $L$. Asymptotically the
value
$-.001136$ is reached, missing the
correct number by about $0.1\%$.}
\label{BRIII}
\end{center}
\end{figure}%
It produces reasonable results also in higher orders,
as is seen in Fig. \ref{BRII}. The
approximations appear
to converge rapidly. But
 the
limit
does not coincide with the
known exact value, although it happens to lie numerically quite close.
 Extrapolating the successive approximations
by an extremely accurate fit to
the
analytically known large-order
behavior \cite{strong} with a function
$b^{(L)}_{0,{\rm plateau}}(\omega=0.843)=A+B\,L^{-\kappa}$,
we find  convergence to $A=1-0.001136$,
which misses the
correct limit $A=1$.
The other two parameters are fitted best by $B=-0.002495$ and
$\kappa=0.922347$ (see
Fig.~\ref{BRIII}).

We may easily convince ourselves by numerical analysis that the
error
in the
limiting value is
indeed linked to the
failure of the
strong-coupling behavior
(\ref{BRAA3})
to have the
power structure
(\ref{STRONG}).
For this purpose
we change
the
function $f(x)$ in equation (\ref{BRAA1.1}) slightly
into $f(x) \to \tilde f(x)=f(x)+1$, which makes the
integrals for $\tilde b_m$ in (\ref{BRAA2}) convergent.
The exact limiting value $1$ of $\tilde \Delta$ remaines unchanged,
but $\bar b_0^{(L)}$
acquires
now the
correct strong-coupling power structure
(\ref{STRONG}).
For this reason,
 we can easily verify
that the
application of
variational theory with a dynamical determination of $ \omega $
yields
the
correct strong-coupling limit $1$
with the
exponentially  fast convergence of the successive approximations
for $L\rightarrow \infty$ like
  $\bar b_0^{(L)} \approx 1-\exp{(-1.909-1.168~L)}$.

In the
next section we are going to point out, that an escape
to complex zeros
which BR propose to
remedy the problems
of the $ \delta $-expansion is really of no help.
\section{The Myth of Complex Zeros and
 Families}
It has been claimed
\cite{BelGarNev} and
repeatedly quoted \cite{RULES}, that the
study of the
anharmonic oscillator in quantum mechanics suggests
the
use of complex extrema to optimize the
$ \delta $-expansion.
In particular, the
use
 of so-called {\em families\/}
of optimal candidates for the
variational parameter
$z$  has been suggested.
We are now going to show, that following these suggestions
one obtains bad  resummation results for the
anharmonic oscillator.
Thus we expect such procedures to lead to even worse results
in field-theoretic applications.

In quantum mechanical applications there
are
 no
 anomalous dimensions
in the
strong-coupling behavior of
the energy eigenvalues.
The
growth  parameters $\alpha$ and
 $\omega$ can be directly read off from
the Schr\"odinger equation;
they are
 $\alpha=1/3$ and
 $\omega=2/3$  for the
anharmonic oscillator
(see
Appendix A).
The
variational perturbation theory is applicable
for all couplings strengths
$g$
as long as $b_0^{(L)}(z)$ becomes stationary for a certain value of $z$.
For higher orders $L$
 it must
exhibit a well-developed
plateau. Within the
range of the
plateau, various derivatives of $b_0^{(L)}(z)$ with respect to $z$ will vanish. In addition there will be complex zeros with
small imaginary parts clustering around the
plateau. They are, however, of limited
use for designing an automatized
computer program
for localizing
 the
position of the
plateau. The study of several examples shows that plotting $b_0^{(L)}(z)$ for various values of $\alpha$
and
 $\omega$ and
 judging visually  the
plateau is by far the
safest method, showing immediately
  which values of $\alpha$
and
 $\omega$
lead to  a    well-shaped
plateau.

\begin{figure}[htp!]
\begin{center}
\setlength{\unitlength}{1cm}
\begin{picture}(19,7)
\scalebox{1.}[1.]{\includegraphics*{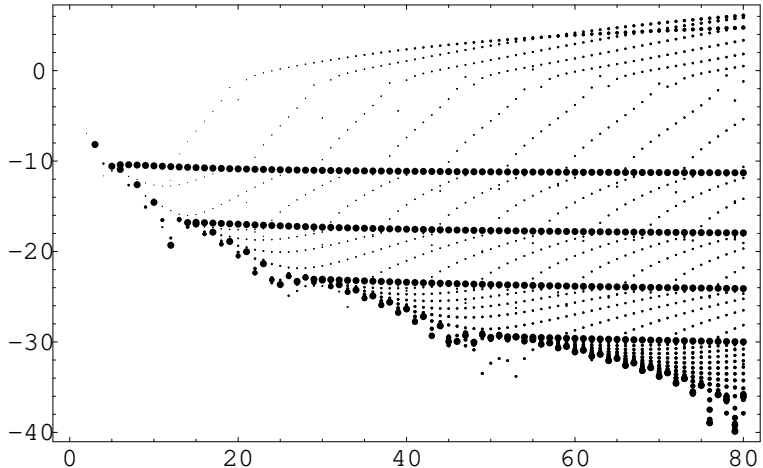}}
\end{picture}
\caption[FigI]{Logarithmic error of the
leading strong-coupling coefficient $b_0^{(L)}$ of the
ground state energy of the
anharmonic oscillator with $x^4$ potential. The errors are plotted over the
order $L$ of the
variational perturbation expansion. At each order, all zeros of the
first derivative have been exploited. Only the
real parts of the
complex roots have been used
to evaluate $b_0^{(L)}$.
The fat points show
results from real zeros, the
smaller points
those
 from complex zeros,
 sizedecreasing
with
distance from real axis.
}
\label{FigI}
\end{center}
\end{figure}

Let us review briefly the
properties
 of the
results obtained from
real and
 complex zeros of $\partial_z b_0^{(L)}(z)$ for the
anharmonic oscillator.
In Fig. \ref{FigI}, the
logarithmic error of $b_0^{(L)}$ is plotted
versus the
order $L$. At each order, all zeros of the
first derivative
are
exploited. To test the rule
suggested in
 \cite{BelGarNev}, only the
real parts of the
complex roots have been used to evaluate $b_0^{(L)}$. The
fat points represent the
results of real zeros, the
thin points stem from the
real parts of complex zeros. It is readily seen that
the
real zeros give the
better result. Only by chance may a complex zero yield a
smaller error.
Unfortunately, there
is no rule to detect these accidental events.
Most complex zeros produce  large errors.

\begin{figure}[htp!]
\begin{center}
\setlength{\unitlength}{1cm}
\begin{picture}(19,11)
\scalebox{1.05}[1.05]{\includegraphics*{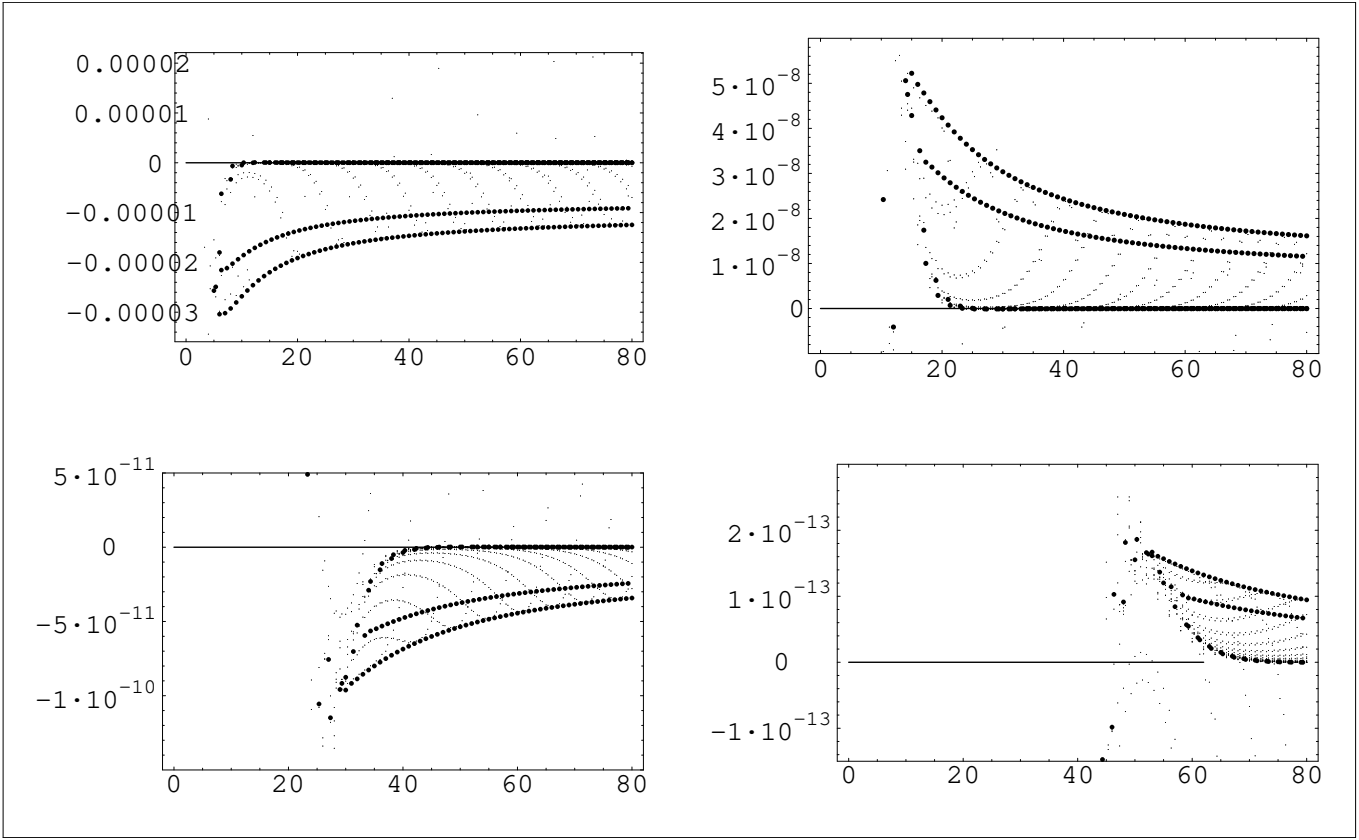}}
\end{picture}
\caption[FgII]{Deviation of the
coefficient $b_0^{(L)}$ from the
exact value is shown as a function of perturbative order $L$ on a linear scale. As before, fat
 dots represent real zeros. In addition to Fig.~\ref{FigI}, the
results obtained from zeros of the
second derivative of $b_0^{(L)}$ are shown.
They give rise to own families with smaller errors by about 30\%.
At $N=6$, the upper left plot shows
 the start of two families
belonging to the first and second
derivative of $b_0^{(L)}$, respectively.
The deviations of both families
are negative.
 On the upper
right-hand figure, an enlargement visualizes the
next two families starting at $N=15$. Their deviations are
positive. The
bottom row shows two more enlargements
of
families starting at $N=30$ and
 $N=53$, respectively. The deviations  alternate again
in sign.
}
\label{FgII}
\end{center}
\end{figure}

We observe the
existence of families described in detail in the
textbook \cite{Hagen} and
 rediscovered in Ref.~\cite{BelGarNev}. These families start at about $N=6,~15,~30,~53$,
 respectively. But each family fails to converge to the
correct result. Only a
sequence of selected members
in each family
leads to an exponential convergence.
 Consecutive families alternate
around the
correct result, as can be seen more clearly in a plot of the
deviations of $b_0^{(L)}$
 from their $L\rightarrow  \infty$ -limit in Fig.~\ref{FgII}, where values derived from the
zeros of the
second derivative of $b_0^{(L)}$ have been included.
These give rise to accompanying families of similar behavior,
deviating with the same sign pattern from the exact result,
but lying closer to the
correct result by about 30\%.
\section{Temperature Shift for $N=2$ Revisited}
Much attention has been paid to a field theoretic model with
$O(2)$-symmetry \cite{BraatenBE,Ramos,Hagen2} to calculate in a realistic context the
coefficient $c_1$, which enters into the
temperature shift of the
Bose-Einstein condensation parametrized as:
\begin{align}
\label{Neq2A}
\frac{\Delta T_c}{T_c^{(0)}}=c_1\,a\,n^{1/3} \,.
\end{align}
Presently, five coefficients of the
relevant perturbation expansion are known
for the
weak-coupling expansion   \cite{BraatenBE,Ramos,Hagen2}
\begin{align}
\label{Neq2B}
F(x)=\sum_{n=-1}^3 a_n\,x^n \,,
\end{align}
whose asymptotic value for $x \to \infty$
coincides with $c_1$:
$c_1=F^*\equiv \lim_{x \to \infty} F(x) \,.$
The known coefficients are
 $a_{-1}=-13.9707$, $a_0=0$, $a_1=-0.446572$, $a_2=0.264412$, $a_3=-0.199$.

We would like to offer an alternative
resummation result for this series to that in Ref.~\cite{Hagen2}. It is
based
on considering
 the
function $x\,F(x)$  containing no negative powers of $x$.
The desired number
 $c_1$ is the the
leading coefficient $b_0$ of the
strong-coupling expansion
\begin{align}
\label{Neq2D}
F(x)=x\left(c_1+\sum_{n=1} b_n x^{-\omega\,n}\right) \,.
\end{align}
The
result for
$c_1$ should be unaffected
by this modification of the function,
 and given by the optimized $L$th-order approximations
\begin{align}
\label{Neq2E}
c_1^{(L)}(z,\omega)=\sum_{l=0}^La_l\, z^{l-1} \binom{L-l+(l-1)/\omega}{L-l}\,.
\end{align}
For the available orders $L\leq 4$,
 this set of functions is now inspected for
plateaus. For $L<3$ there is none. For $L=3$ and
 $L=4$, a
plateau can be identified unambiguously as the
only horizontal turning point
solving  simultaneously
 $\partial_z c_1^{(L)}(z,\omega)=0$ and
 $\partial_{z}^2 c_1^{(L)}(z,\omega)=0$.
The results are
\begin{align}
\label{c1}
L &= 3  &
z^{(3)} &= 1.089  &
\omega^{(3)} &= 1.071  &
c_1^{(3)} &= 0.940\\
L &= 4  &
z^{(4)} &= 2.057  &
\omega^{(4)} &= 0.571  &
c_1^{(4)} &= 1.282
\end{align}
Given only two approximations
for $c_1$ it is unrealistic to attempt
an extrapolation to $L \to \infty$, as done with
another selection rule of optima in Ref.~\cite{Hagen2},
but it is interesting to note that the
value of the
coefficient for the
temperature shift $c_1^{(4)}=1.282$ is in excellent agreement with the
latest Monte Carlo result of $c_1 \approx 1.30$ \cite{C1MC}.
\section{Renormalization Group and
 Variational Perturbation}
The most convincing evidence for the power of
 the field-theoretic variational perturbation
theory with anomalous dimensions
comes from applications to
critical exponents in $4- \epsilon $ dimensions \cite{epsilon,Verena}.
The results obtained turn out to be
immediately resummed expressions
of  the $ \epsilon $-expansions, which can be recovered as
a Taylor series.
The renormalization group function $\beta(g)$
is obtained from
the weak-coupling expansion of the
renormalized coupling constant $g$ in terms of the
bare coupling constant ${g_B}$ \cite{RenGroup,Verena}:
\begin{align}
\label{beta}
\beta(g,\epsilon)=-\epsilon~g~\frac{d~\log g({g_B},\epsilon)}{d~\log~{g_B}}=-\epsilon~g~\Big[\frac{d~\log {g_B}(g,\epsilon)}{d~\log~g}\Big]^{-1}~.
\end{align}
Due to renormalizability, $\beta(g)$ necessarily has the
form
\begin{align}
\label{beta2}
\beta(g,\epsilon)=-\epsilon~g+\beta_0(g)~.
\end{align}
Perturbation theory
with minimal subtractions yields the
weak-coupling expansion:
\begin{align}
\label{g-Coupling}
g={g_B}+\sum_{k=1}^\infty f_k({g_B})~\epsilon^{-k}~,
\end{align}
where $f_k({g_B})$ possesses an expansion in powers of ${g_B}$, starting with ${g_B}^{k+1}$.
 By suitably normalizing $g$ and
 ${g_B}$, the
leading coefficient of $f_1$ can be made equal to minus one: $f_1({g_B}) =-{g_B}^2+O({g_B})^3$.
The function $\beta_0(g)$ can be expressed in terms of the
residue $f_1({g_B})$ of the
$\epsilon$-pole in equation(\ref{g-Coupling}) alone:
\begin{align}
\label{beta0}
\beta_0(g)=f_1(g)-g\,f_1'(g)~.
\end{align}
Recall the standard proof for this
based on combining
Eqs.~(\ref{beta}) and
 (\ref{beta2}) to
\begin{align}
\label{bx}
\beta_0(g)=\epsilon\,g-\epsilon\,{g_B}~\partial_{g_B}\, g({g_B},\epsilon)~,
\end{align}
which becomes, after inserting equation (\ref{g-Coupling}):
\begin{align}
\label{bxx}
\beta_0\Big[{g_B}+\sum_{k=1}^\infty f_k({g_B})~\epsilon^{-k}\Big]
=\epsilon\,\sum_{k=1}^\infty \Big[f_k({g_B})-{g_B} f_k'({g_B})\Big] \epsilon^{-k}~.
\end{align}
The
limit $\epsilon \to \infty$ leads directly to the property (\ref{beta0}).

Another well-known fact is
that
all the
functions $f_k({g_B})$ for $k>1$ can be expressed in terms of the
residues $f_1({g_B})$ only
\cite{Verena}. Indeed,
taking the derivatives of $\beta_0(g)$ in equation (\ref{bx})
with respect to ${g_B}$ and
 $\epsilon$:
\begin{align}
\label{by}
\beta_0'(g)~\partial_{g_B} g   &= -\epsilon \,{g_B} ~\partial_{g_B}^2\,g,\\
\beta_0'(g)~\partial_\epsilon  g &=g+\epsilon ~\partial_\epsilon g-{g_B}~\partial_{g_B} g-\epsilon \, {g_B}~\partial_{g_B} \partial_\epsilon g~,
\end{align}
eliminating $\beta_0'(g)$ between these two equations,
 and
 inserting the
expansion (\ref{g-Coupling}), we obtain order by order in $1/\epsilon$ a recursive set of
differential equations for the
functions $f_k({g_B})$ with $k>1$, which are
 power series in $g_B$. If we now expand
\begin{align}
\label{power}
f_1({g_B}) &= -{g_B}^2+\sum_{j=3}^\infty \gamma_j {g_B}^j~, &
f_k({g_B}) &= \sum_{j=k+1}^\infty \gamma_{k,j} {g_B}^j ,
\end{align}
a solution is readily found, beginning with
\begin{align}
\label{sol1}
\gamma_{k, k+1} &=(-1)^k,
& \gamma_{2, 4} &= -\frac{8}{3}\, \gamma_3,
&\gamma_{2, 5} &= \frac{3}{2}\, \gamma_3^2-\frac{7}{2}\,\gamma_4,\\
\gamma_{3, 5} &= \frac{29}{6}\,\gamma_3,
&\gamma_{2, 6} &= \frac{18}{5}\, \gamma_3\,\gamma_4- \frac{22}{5}\, \gamma_5,
& \gamma_{3, 6} &= -\frac{32}{5}\, \gamma_3^2+ \frac{39}{5}\, \gamma_4,\\
\gamma_{4, 6} &= -\frac{37}{5}\, \gamma_3,
& \gamma_{2, 7} &= 2\, \gamma_4^2 + \frac{13}{3}\, \gamma_3\, \gamma_5 - \frac{16}{3}\,\gamma_6,
& \gamma_{3, 7} &= \frac{5}{2}\,\gamma_3^3 - \frac{551}{30}\,\gamma_3\,\gamma_4 + \frac{59}{5}\,\gamma_5 ,\\
\gamma_{4, 7} &= \frac{751}{45}\, \gamma_3^2 - \frac{141}{10}\, \gamma_4,
& \gamma_{5, 7} &= \frac{103}{10}\,\gamma_3.
\label{sol2}
\end{align}
In the renormalization group approach, a
 fixed point $g^* \neq 0$ is determined by the
zero of the
$\beta$-function: $\beta(g^*)=0$. The Wegner exponent $\omega$
 governing the
approach to scaling
 is given by the
slope
at the
fixed point: $\omega=\beta'(g^*)$. The two quantities
have $ \epsilon $-expansions
\begin{align}
\label{gStar}
g^* &=\sum_{j=1}^\infty \alpha_j \epsilon^j ,&
\omega &=\sum_{j=1}^\infty \omega_j \epsilon^j~.
\end{align}
The coefficients $\alpha_j$  and
 $\omega_j$ are determined from  the
residues $\gamma_j$ as:
\begin{align}
\label{alpha}
\alpha_1 &= 1,&
\alpha_2 &= 2~\gamma_3, \hspace{2.cm}
\alpha_3 = 8~\gamma_3^2+3~\gamma_4\\
\alpha_4 &= 40~\gamma_3^3+30~\gamma_3~\gamma_4 + 4~\gamma_5  ,&
\alpha_5 &= 224~\gamma_3^4 + 252~\gamma_3^2~\gamma_4 + 27 ~\gamma_4^2~+ 48~\gamma_3~\gamma_5 +5~\gamma_6,
\label{alpha2}
\end{align}
and
\begin{align}
\label{omega}
\omega_1 &= 1, &
\omega_2 &= -2~\gamma_3, &
\omega_3 &= -8~\gamma_3^2-6~\gamma_4 ,&
\omega_4 &= -40~\gamma_3^3 - 48~\gamma_3~\gamma_4 - 12~\gamma_5.
\end{align}
We can now convince ourselves
that precisely
 the
same results can be derived from
 variational perturbation theory applied to
the weak-coupling expansion (\ref{g-Coupling})
(and as shown in \cite{without}
from the expansion of any other critical exponent).
  We determine $\omega$ dynamically solving Eq.~(\ref{LOG}).
We insert for $\omega$ an unknown
$\epsilon$-expansion of
the form (\ref{gStar}). The variational parameter $z$
is then adjusted to make (\ref{LOG}) stationary.
Then, since for $\epsilon \to 0$ the weak-coupling coefficients
of $g(g_B)$ in the expansion (\ref{g-Coupling}) behave like
 $ \sim \epsilon^{1-l}$,
 $z$ has to scale with $\epsilon$, so that we may put
 $z= \zeta _1\epsilon+\zeta_2~\epsilon^2+\zeta_3 \epsilon ^3+O(\epsilon^4)$, and
 solve equations (\ref{LOG}) and
 (\ref{LOG1}) for each perturbative order $L$,
 order by order in $\epsilon$. This leads to a rapidly increasing number of non-linear and
 not even independent equations for the
unknown $\zeta_l$ and
 $\omega_l$, some depending also on the
order $L$.

Despite these possible complications, the
solutions turn out to be well structured and
 easily obtained. At each $L$
to lowest order in  $\epsilon$,  the
term independent of $\epsilon$ in (\ref{LOG}) and
 the
coefficient of $\epsilon^{-1}$ in (\ref{LOG1}) demand
 that $\zeta_1=1$. In addition, they require
$\gamma_{k,k+1}=(-1)^k$ for some $k$,
in agreement with Eqs.~(\ref{sol1}). Such conditions
imposed
 on
$\gamma_{k,l}$ can, of course, not depend on the
order $L$, but must be enforced in general. Raising  the
order of $\epsilon$ in (\ref{LOG}) and
 (\ref{LOG1}), and
 imposing $\zeta_1=1$ as well as the
conditions already established for the
$\gamma_{k,l}$, all dependences on the
$\omega_k$ and
 $\zeta_k$ disappear,
and we are left with conditions on
 $\gamma_{k,l}$ alone, which reproduce exactly the
relations (\ref{sol1}) through (\ref{sol2}). This shows, that the
variational perturbation method is completely
compatible
the
 well-known $\epsilon$-expansions,
 if the
input divergent series has a structure
 satisfying the
renormalization group equation (\ref{beta2}).

After having reproduced
$\gamma_{k,l}$, there are further equations to be solved. Going to the
next higher order in  $\epsilon$, either for
 (\ref{LOG}) or for (\ref{LOG1}), gives a relation
 involving exactly one of the
expansion coefficients of $\epsilon /\omega$,
 which are simply related to the
coefficients $\omega_l$ of $ \omega $.
In this way, the
renormalization group results of (\ref{omega}) are exactly reproduced.
These solutions are stable in the
sense, that with increasing order $L$, the
expansion coefficients $\omega_l$ for $l<L$ remain unchanged.
This proves, that the
variational method produces
the same $ \epsilon $-expansions
of all critical exponents
as renormalization group theory.
 At the
same time this implies
 that the standard $ \delta $-expansion
which does not allow for
the
 anomalous dimension $ \omega $ is bound to fail.


It is noteworthy, that several other conditions are
automatically satisfied up to some order $\epsilon^{L}$, $\epsilon^{L-1}$,
 or $\epsilon^{L-2}$, respectively.
Among them is the
variationally transcribed second logarithmic derivative of the
weak-coupling series and
 the
derivative thereof:
\begin{align}
\label{LOGLOG}
\sum_{l=0}^L\; h_l\;z^l \binom{L-l+l/\omega}{L-l} &= -1-\omega,\\
\sum_{l=1}^L\; h_l\;l\,z^{l-1} \binom{L-l+l/\omega}{L-l} &=0 \,,
\end{align}
where the
$h_l$ are the
expansion coefficients of
\begin{align}
\label{LOGLOG2}
\frac{{g_B}~g''({g_B})}{ g'({g_B})}\,.
\end{align}
Of some computational benefit is the
observation, that with the
same accuracy in  $\epsilon$ the
first and
 second derivatives of the
variational series (\ref{B0}) themselves  vanish (here for $\alpha=0$). This means, that the
function has a
flat
plateau. For a typical field-theoretic application with only a
few known perturbation coefficients, the
plateau is easily found by inspection. Therefore, if the
model possesses a well-behaved $\beta$-function satisfying equation (\ref{beta2}), we expect a reliable result for the
anomalous dimension $\omega$ if it is chosen such as to produce an acceptable
plateau. The ordinate of the
plateau is the
most promising variational perturbative value for the
quantity analyzed to the
respective order.

\section{Conclusion}
Summarizing
this paper
 we have learned
that
the so-called  $ \delta $-expansion is
inapplicable to
quantum field theory, since it does not
account for the
 Wegner exponent $ \omega $ of approach to the strong-coupling limit.
Only the field-theoretic variational perturbation theory
yields correct results
by
incorporating $ \omega $ in an essential way.

\section{Appendix A}
Here we review briefly
 how the
strong-coupling parameters
$ \alpha $ and $  \omega $
in (\ref{STRONG})
and the
variational equation (\ref{B0})
for the leading strong-coupling coefficient
are found
for the anharmonic oscillator
with the
Schr\"odinger equation
in natural units
\begin{align}
\label{A1}
-\frac{1}{2}\Psi''+\frac{x^2}{2}\Psi+g\,x^{2\kappa}\,\Psi=E\,\Psi \,.
\end{align}
We rescale the
space coordinate $x$ so
 that the
potential becomes
\begin{align}
\label{A2}
V(x)=\frac{1}{2}g^{-2/(\kappa+1)}x^2+x^{2\kappa}\,,
\end{align}
any eigenvalue
has
the obvious
strong-coupling expansion
\begin{align}
\label{A3}
E=g^{1/(\kappa+1)} \sum_{l=0}^\infty b_l\big(g^{-2/(\kappa+1)}\big)^l\,,
\end{align}
where
$b_l$ are the
strong-coupling coefficients. The aim is to determine them from the
known weak-coupling coefficients $a_n$
of the  divergent perturbation expansion:
\begin{align}
\label{A4}
E=\sum_{l=0}^\infty a_lg^l\,.
\end{align}
The solution of this problem
comes
 from physical intuition,
suggesting
that
the perturbation expansion
should be performed
around an effective harmonic potential  $ \Omega ^2x^2/2$,
whose frequency is different from the bare value 1/2
in (\ref{A1}), depending on $g$ and the order $L$  of truncation of (\ref{A3}).
Thereafter only the
difference between the
anharmonic part and
 the
effective harmonic part is to be treated by perturbation methods. The
trial frequency
 $\Omega$ of the
effective potential can be fixed later by the
consideration, that the
resulting quantity of interest should
be as independent as possible
of $\Omega$, according to
 the
principle of minimal sensitivity.
With the
harmonic trial potential
$V_{ \Omega }^{{(0)}}=\Omega^2x^2/2$,
the
interaction potential
(\ref{@varint})
reads
 $
V_{ \Omega }^{\rm int}= \delta \left[ g\,x^{2\kappa}-(\Omega^2-1)x^2/2\right] $.
The parameter $ \delta $ organizes the
reexpansion
and
 is set equal to $1$ at the
end.
The expansion proceeds from the
rescaled
Schroedinger equation (\ref{A1}):
\begin{align}
\label{A5}
-\frac{1}{2}\Psi''+\frac{x^2}{2}\Psi+\frac{\delta\,g\,x^{2\kappa}\,\Psi}{\beta^{N+1}}=\frac{E}{\beta}\,\Psi \,,
\end{align}
where $\beta=\sqrt{\Omega^2-\delta( \Omega ^2-1)}$.
To order $L$, the
energy has the
reexpansion
\begin{align}
\label{A6}
E^{(L)}(\Omega,g)=\beta\,\sum_{l=0}^L a_l^{(i)}\Big(\frac{\delta\,g}{\beta^{\kappa+1}}\Big)^l.
\end{align}
with
the
well-known weak-coupling expansion coefficients
as defined in equation (\ref{A4}). The
strong-coupling behavior  (\ref{A3}) suggests
 changing the
variational parameter from $\Omega$ to
$z:=\frac{g}{\Omega^{\kappa+1}}.$
In the
limit $g \to \infty$ we obtain
the reexpansion which must be optimized in $z$:
\begin{align}
\label{A8}
E^{(L)}(z)= g^\alpha\;\sum_{l=0}^L\; a_l^{(i)}\;z^{l-\alpha} \binom{L-l+(l-\alpha)/\omega}{L-l}
\end{align}
where $\omega=2/(\kappa+1)$ and $\alpha=1/(\kappa+1)$. For the
leading coefficient of the
strong coupling expansion of the
ground state energy, Eq.(\ref{A8})
leads directly
to the variational equation (\ref{B0}).

\section{Appendix B}
In order to gain further insight into the
working of the variational resummation procedure, we apply it to the simple
test
function
\begin{align}
\label{E1}
f(x)=(1+x)^\alpha=x^\alpha \Big(1+\frac{1}{x}\Big)^\alpha
\end{align}
with
 weak coupling coefficients $a_n=\binom{\alpha}{n}$
and
a leading strong-coupling behavior $f\sim x^\alpha(1+{\alpha}/{x}+\dots)$,
so that
 $b_0=1$.
Inserting this information into equation (\ref{B0}), we obtain
the variational
leading coefficient to $L$th order:
\begin{align}
\label{E2}
b_0^{(L)}(z)=\sum_{l=0}^L\binom{\alpha}{l}\binom{L-\alpha}{L-l}\,z^{l-\alpha},
\end{align}
which is easily transformed into the
expression:
\begin{align}
\label{E3}
b_0^{(L)}(z)=\binom{\alpha}{L}\sum_{l=0}^L\binom{L}{l}\,\frac{L-\alpha}{l-\alpha}\,(-1)^{L+l}z^{l-\alpha}
\end{align}
Determining the
variational parameter $z$ according to the
principle of minimal sensitivity requires a well developed
plateau of $b_0^{(L)}$ as a function of $z$.
For the
simple test function,
 the derivative $\partial_z b_0^{(L)}(z)$ can be obtained in the closed form:
\begin{align}
\label{E4}
\frac{d}{dz}b_0^{(L)}(z) &=(-1)^{L+1}\frac{L-\alpha}{z^{\alpha+1}}\binom{\alpha}{L}\sum_{l=0}^L(-z)^l\binom{L}{l}\\
 &=(-1)^{L+1}(L-\alpha)\binom{\alpha}{L}\frac{(1-z)^L}{z^{\alpha+1}}~.
\end{align}
This exhibits a flat plateau around
$z=1$ if the order $L$ is much larger than $\alpha$. An equally flat
 plateau is found for $b_0^{(L)}(z)$.
The
value of the
leading strong coupling coefficient $b_0^{(L)}$ at the
plateau is
\begin{align}
\label{E5}
b_0^{(L)}(1)=\binom{\alpha}{L}\sum_{l=0}^L\binom{L}{l}\,\frac{L-\alpha}{l-\alpha}\,(-1)^{L+l}=1,
\end{align}
in perfect agreement with the exact result, thus
confirming the
applicability of the
resummation scheme for this class of problems.
\end{document}